\documentclass[letterpaper,twocolumn,10pt]{article}
 \usepackage{tikz}
\usepackage{amsmath}

\usepackage{filecontents}
\usepackage{hyperref} 
\usepackage{enumitem}

\begin{document}

\date{}

\title{\Large \bf Fast Raft: Optimizations to the Raft Consensus Protocol}

\author{
{\rm Bryan SebaRaj}\\
Yale University
\and
{\rm Anton Melnychuk}\\
Yale University
} 

\maketitle

\begin{abstract}
We present the open-source implementation and evaluation of Fast Raft, a new consensus algorithm by Castiglia, Goldberg, and Paterson designed for dynamic networks. Fast Raft, is a variation on the Raft consensus algorithm that reduces the average number of message rounds to commit a log entry in typical operation\cite{fastraft}. Fast Raft is ideal for fast-paced distributed systems where membership changes over time and where sites must reach consensus quickly. We develop and present experimental evaluations comparing our implementation to the Raft algorithm deployed on AWS (EKS). These evaluations draw some parallels with the Fast Raft paper and highlight the hidden challenges we encountered.
\end{abstract}

\section{Introduction}

State machine replication serves as a foundational processes in providing fault tolerance and increasing availability across distributed systems. Consensus algorithms provide a fast, fault-tolerant method to achieve such an agreement on the order of updates, with consensus being used across a multitude of distributed systems. Modern distributed systems are increasingly large-scale, globally distributed, and highly dynamic. 



Traditional consensus algorithms such as Raft \cite{raft}. and Paxos \cite{paxos} were not designed with these challenges in mind. Exploring newer approaches, such as Fast Raft or Fast Paxos, is therefore crucial. Without an open-source implementation of Fast Raft from Castiglia, Goldberg, and Paterson, we set out to implement and test the algorithm, analyzing its real-world performance and the associated challenges.

Our implementation of Fast Raft builds upon the foundation of the network-simulated Raft consensus algorithm provided as an assignment for the Building Distributed Systems course at Yale. It closely follows the Fast Raft algorithm described in Sections IV and V of the original paper \cite{fastraft} and aims to accomplish the following milestones:
\begin{enumerate}[nosep]
    \item Containerize a deployable Raft node with gRPC.
    \item Extend the implementation to Fast Raft.
    \item Deploy to EKS and test by simulating network failures.
\end{enumerate}

We aim to demonstrate how Fast Raft outperforms classical Raft in handling network disruptions, focusing on throughput and latency on achieving log consensus.







\section{Design and Implementation}

\subsection{Buiding the Raft Node}

In order to build deployable Raft/FastRaft nodes, we first started with lab 3 (Raft) from CPSC 426. Since a raft cluster is typically deployed across multiple hardware nodes, we needed a robust networking protocol which facilitated intra-node communication. In order to reduce message complexity and overhead, we elected to use gRPC over JSON. We configured each node as a gRPC server, establishing gRPC clients with all other raft nodes within their respective clusters.We implemented and modified essential server functions such as \texttt{AddReplica} for bootstrapping, \texttt{ApplyCommand} and \texttt{GetLogs} for correctness and load testing, \texttt{AppendEntries} and \texttt{RequestVote} for the election process, and \texttt{CommitOperation} and \texttt{ForwardOperation} for replication. In addition, in order to properly handle \texttt{performCommit()} from the non-leader sites, we transferred the command over gRPC to the current raft cluster leader, and wrote an executable function to bootstrap the raft node within its cluster.


\subsection{Fast Raft}

We extend our Raft implementation with Fast Raft optimizations, summarized here for brevity. Fast Raft offers two paths for log commitment: a “fast track” and the traditional “classic track.” In the fast track, when a node proposes a new entry $e$ for log index $i$, it sends the entry directly to all nodes designated as “self-approved.” Upon receiving $e$, these nodes each process and tentatively insert it into their logs at index $i$. The leader then collects votes for the entry and finalizes it if $\lceil\frac{3M}{4}\rceil$ nodes support it, ensuring consistency while reducing leader-dependency (and it's resulting bottleneck).

This fast-track approach requires careful handling of potential log inconsistencies, since sites may encounter overlapping proposals or miss consecutive indexes, since the log is now over writable and not an append-only log. If there are insufficient votes for a proposal, or if concurrent proposals cause conflicts, Fast Raft gracefully reverts to the classic Raft algorithm, incurring additional overhead. In environments where packet loss is high, the classic track may still outperform the fast track, but in lower-loss settings, Fast Raft’s parallel approach can significantly improve throughput and reduce commit latency.

Overall, Fast Raft’s design remains faithful to Raft’s core principles of safety, liveness, and (relative) simplicity, while introducing a performance-oriented mechanism for proposing log entries. By limiting reliance on a single leader and exploiting parallel vote replication, Fast Raft can better handle scenarios with bursts of writes or less-than-ideal network conditions, as long as proposals remain largely non-conflicting.

\subsection{Deployment to AWS}

The Raft and Fast Raft clusters, alongside a load-tester, where deployed to AWS (us-east-2) across three availability zones via Terraform. Each cluster was given its own EKS nodegroup, consisting of 3 t2.micro (Linux/AMD64) EC2 instances, with each node in a different availability zone. The clusters were deployed as a stateful set of three privileged (see section 3), headless (to bypass kube-proxy's loadbalancing step to marginally reduce overhead), anti-affinity pods, such that each pod had its own kubernetes node (t2.micro instance) in order to simulate the desired availability and durability that state machine replication was intended to provide. The stateful sets also provided the Raft and Fast Raft pods with constant IPs, which made reconnection to nodes after crash failures trivial over gRPC. The load-tester was deployed as an ephemeral pod with a load-testing executable. 

\section{Experiments}

\subsection{Correctness}

In order to simulate a ``lossy'' network, we used the Linux \texttt{tc} util. We performed numerous tests to mirror the unit tests provided in lab 3 from CPSC 426 to test the correctness of the Raft and Fast Raft clusters under conditions of random packet loss and complete packet loss/network outage and network delays to and from various nodes. In addition, crash failures of individual Raft and Fast Raft nodes were simulated by killing the respective stateful set pod deployed on EKS. Sample inputs to the Raft and Fast Raft nodes were simulated from the load-tester pod within the EKS cluster, running on a separate nodegroup/t2.micro instance, and the logs of the nodes of each cluster were compared to confirm the correctness of the state machine replication algorithms.  

\subsection{Performance}

To measure latency across the Raft/Fast Raft clusters, the load-tester was used to send bursty simulated workloads, with the average latency measured across the nodes under various random packet loss conditions. All tests yielded a 0\% failure rate, and their average latencies were plotted in Figure 1. As demonstrated and similar to the results from the original Fast Raft paper, Fast Raft yields a marginally reduced latency under low packet loss conditions, while increasing in latency after packet loss increases past 4\% due to the it's additional overhead failures before falling back to the classical track. 

However, under real world networking scenarios, where the packet loss is often well under 2\%, the demonstrated increases in performance from Fast Raft is relevant. This ultimately confirms the improvements yielded by the original Fast Raft paper while maintaining the safety and liveness requirements of a consensus algorithm for state machine replication.  

\begin{figure}[h]
    \centering
    \includegraphics[scale=0.3]{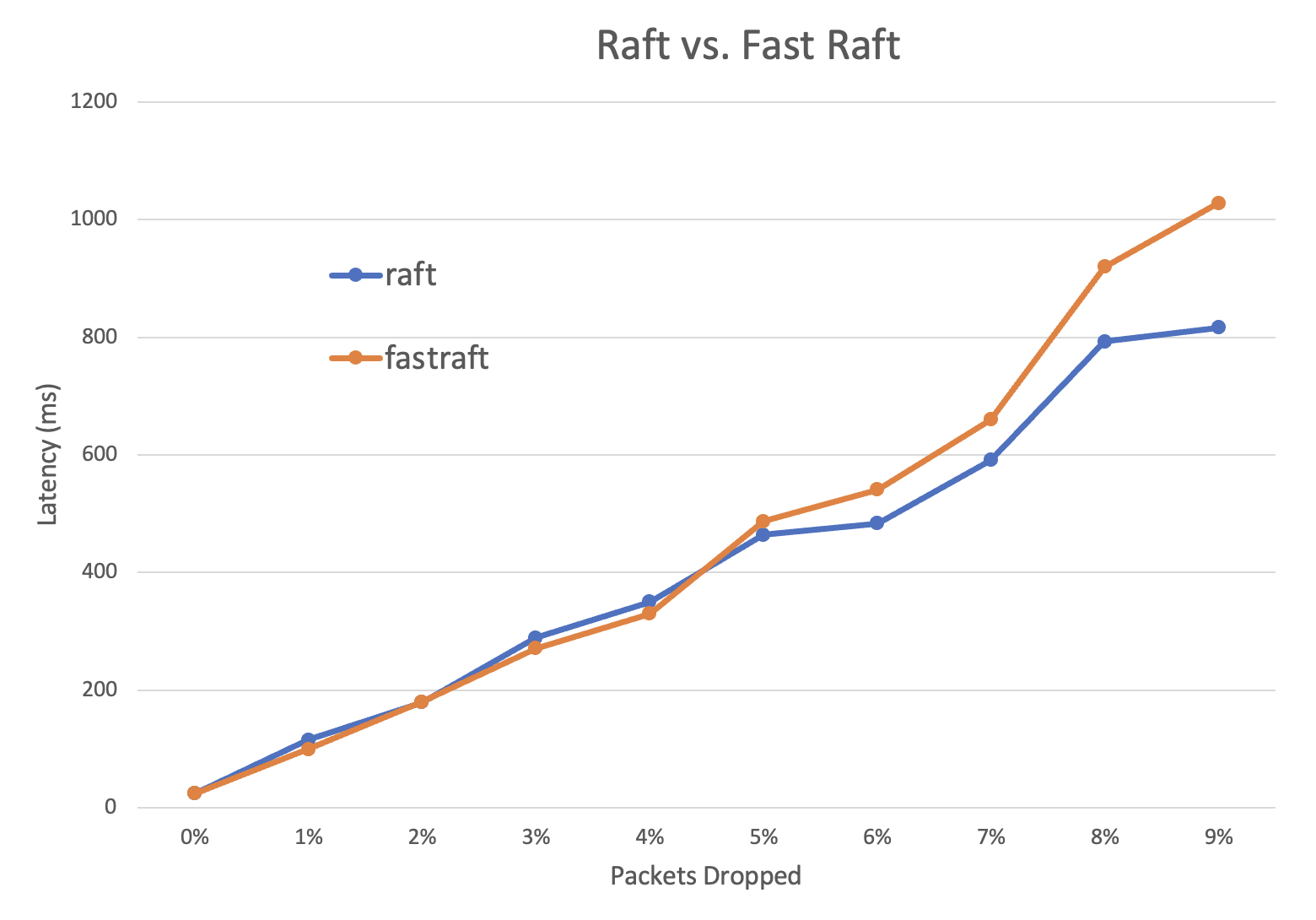}
    \caption{Latency of Raft and Fast Raft clusters on EKS at various levels of random packet loss.}
    \label{fig:evaluations}
\end{figure}

\section{Related Work}

Numerous consensus protocols have been proposed to tackle the challenges of consistency and fault tolerance in distributed systems. Paxos \cite{raft}, widely regarded as a foundational algorithm in this space, underpins many production systems due to its well-understood guarantees and broad adoption. Later iterations, such as Fast Paxos \cite{fastpaxos} and Multi-Paxos \cite{multipaxos}, optimized the original protocol by batching operations, reducing message complexity, and minimizing leadership changes. These optimizations, while effective in raising throughput, come with added implementation and conceptual overhead that can complicate system design.

Raft \cite{raft} emerged as an alternative designed to be more approachable and maintainable. Its emphasis on understandability made it attractive for both research and industrial applications, leading to multiple open-source implementations across various languages. Many production-grade systems adopted Raft to coordinate distributed data management, handle cluster membership changes, and maintain high availability. Although Raft’s leader-based approach simplifies the consensus process, the leader can become a performance bottleneck when faced with high-volume, write-intensive workloads, or geo-distributed deployments.

FastRaft \cite{fastraft} builds upon these ideas, retaining the simplicity and fault-tolerance guarantees of Raft while introducing mechanisms to reduce leader contention and streamline data replication. By drawing inspiration from both FastPaxos, Multi-Paxos, and other Raft-based optimizations, FastRaft uses better parallelization and lightweight leader-election enhancements to address throughput and latency constraints. This approach represents an evolution rather than a departure from prior work, showing that Raft’s design can be feasibly adapted for more demanding environments without sacrificing its readability or proven correctness properties. Our demonstrated performance gains at low levels of packet loss, even though marginal, are easily applicable to a real-world setting, as the percent of packets dropped is often less than 2\%, and as such, shows how the optimizations presented by Fast Raft can improve the performance of a consensus algorithm which underpins many critical infrastructures reliant on Raft state machine replication. 

\section*{Acknowledgments}

We would like to thank Professor Richard Yang, Scott Pruett, and Xiao Shi for their invaluable expertise and advice during this project.





\bibliographystyle{plain} 
\bibliography{references} 

\end{document}